\documentclass[twocolumn,showpacs,aps,prb,amsmath,amssymb,floatfix]{revtex4}
\usepackage{graphicx}
\begin{document}
\title
{Transmission through Quantum Dots: Focus on Phase Lapses}
\author{D. I. Golosov$^1$ and Yuval Gefen$^2$}
\affiliation{
 $^1$Racah Institute of Physics,
the Hebrew University, Jerusalem 91904, Israel\\
 $^2$Dept. of Condensed Matter Physics, Weizmann Institute of Science,
Rehovot 76100, Israel}
\date{\today}
\begin{abstract}
Measurements  of the  transmission phase in transport through a
quantum dot embedded in an Aharonov-Bohm interferometer  show
systematic  sequences of phase lapses  separated by  Coulomb
peaks. Using  a two-level  quantum dot as an example we show that
this phenomenon can be accounted for by the combined effect of
asymmetric  dot-lead coupling and interaction-induced "population
switching" of the levels, rendering this behavior generic.
In addition, we use the notion of spectral shift function to analyze the 
relationship between transmission phase lapses and the Friedel sum rule.
\end{abstract}

\maketitle

In a series of experiments by the Weizmann group, the transmission
phase, $\Theta_{tr}$,  characterizing  transport through a quantum
dot (QD) has been systematically studied
\cite{Yacoby,Schuster,Avinun}, embedding the QD in an
Aharonov-Bohm interferometer\cite{GIA,Ora}.  Arguably the most
intriguing finding of these experiments has been the correlated
behavior of $\Theta_{tr}$ as function of the leads' chemical
potential $\mu$ (or the gate voltage): it appears to undergo a
lapse (phase lapse, PL), seemingly of $-\pi$, between any two
consecutive Coulomb peaks. It is  clear that this effect cannot be
explained within a single-particle framework\cite{BGEW}. Moreover,
in spite of a substantial body of theoretical work (see, {\it
e.g.,} Refs. \onlinecite{YGreview,Hackenbroich,Aharony}), some of which
gained important insight  on the underlying physics, no clear cut
theory-experiment connection has been established as yet.

In the present article we revisit this problem. We do this by
studying a (spinless) two-level QD, attached to two leads. We
account for the
 difference in the  couplings of level 1 and level 2
to the leads ("$1-2$ asymmetry") and, for the first time,  probe
the effect of the (generically expected) asymmetric coupling to
the left and the right leads ("L-R asymmetry"). We find
unexpectedly that these two asymmetries give rise to a
qualitatively new behavior of $\Theta_{tr}(\mu)$, and render the
appearance of PL between consecutive Coulomb peaks generic.
This conclusion is in line with recent renormalisation group results
for a QD with degenerate levels\cite{Marquardt}.

Throughout the  discussion of transmission PLs in the literature, 
much attention was paid to the Friedel sum rule, which, in one dimension, 
relates
the transmission phase to the change of carrier population in the system
(see, {\it e.g.,} Refs. \onlinecite{Anderson,But99}).
Since the latter varies monotonously with the chemical potential (or
gate voltage), one may perceive a contradiction between this sum
rule and the occurrence of PLs.  We revisit this issue in Appendix
\ref{app:shift} and show, in particular, that the correct formulation
of Friedel sum rule in one dimension allows for transmission phase
lapses. 

\begin{figure}
\includegraphics{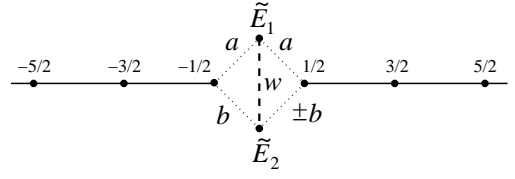}
\caption{\label{fig:schemedot} The model system, composed of a
wire (chain) and a two-level dot, Eqs. (\ref{eq:tunnel}) and
(\ref{eq:intradot}).}
\end{figure}

The minimal model for studying the phase lapse mechanism includes
a two-level QD,
\begin{equation}
{\cal H}_{QD} = (E_1^{(0)}-\mu ) \hat{d}^\dagger_1 \hat{d}_1 + (
E_2^{(0)}-\mu ) \hat{d}^\dagger_2 \hat{d}_2+ U \hat{d}^\dagger_1
\hat{d}^\dagger_2 \hat{d}_2 \hat{d}_1\,.
\label{eq:Hdotgen}
\end{equation}
Here, the operators $\hat{d}_i$ with $i=1,2$ annihilate  electrons
on the two dot sites (with bare energies $E^{(0)}_i$,
$E^{(0)}_2>E^{(0)}_1$). The QD is coupled to the two leads by the
tunnelling term
\begin{eqnarray}
V_T=&&- \frac{1}{2} \hat{d}_1^\dagger \left( a_L \hat{c}_{-1/2}+
a_R \hat{c}_{1/2} \right)-
\nonumber \\
&&-\frac{1}{2} \hat{d}_2^\dagger \left(b_L \hat{c}_{-1/2}+b_R
\hat{c}_{1/2} \right)+ {\rm h. c.}.
\label{eq:tunnel}
\end{eqnarray}
The operators $\hat{c}_j$ (with half-integer $j$) are defined on
the tight-binding sites of the left and right lead (cf. Fig.
\ref{fig:schemedot}).

We begin with summarizing the results of Ref.\onlinecite{SOG} (see also
Refs. \onlinecite{OG97,Weidenmueller,Kim,But99}) in the case when no charging
interaction is present, $U=0$, and the value of $\Theta_{tr}$ is
readily calculated (even for a larger number of dot levels). The
two  transmission peaks then take place near $\mu=E^{(0)}_i$; each
corresponds to a smooth increase of $\Theta_{tr}(\mu)$ by $\pi$
within a chemical potential range proportional to $a_L^2+a_R^2$
for the first dot level, $b_R^2+b_L^2$ for the second one. If the
relative coupling sign, $\sigma \equiv {\rm sign} (a_L a_R b_L
b_R)$, equals $+1$ ({\it same-sign case}), a discontinuous PL of
$\Delta \Theta_{tr}=-\pi$ (transmission zero) arises in the energy
interval between the two transmission peaks, $E_1^{(0)} < \mu
<E_2^{(0)}$. While this would be in qualitative agreement with the
measurements, experimentally there is no way to control the
coupling signs. Indeed, for the relevant case of a random
(chaotic) QD, one expects close to 50 \% of the adjacent pairs of
dot levels to have $\sigma=-1$ ({\it opposite-sign case}), when no
phase lapse occurs between the two corresponding level crossings.
These  observations \cite{SOG} (and hence the difficulty in
accounting for the experimentally observed correlations in
$\Theta_{tr}$) persist even  when interaction is accounted for
(but when $|a_L|=|a_R|=|b_L|=|b_R|$ was assumed).

Following the original idea of Ref. \onlinecite{SI}, the effects of   
"population
switching"  due to a charging interaction U in discrete
 spectrum QDs [Eq. (\ref{eq:Hdotgen})]  were addressed both
theoretically\cite{Baltin,BvOG,YG04,Sindel,Dagotto} and
experimentally\cite{ExpInt}.
If one of the dot levels is
characterized by a stronger coupling to the leads and $U$ is
sufficiently large,  the two level occupancies, $n_i=\langle
\hat{d}^\dagger_i \hat{d}_i \rangle$  show non-monotonic
dependence on  $\mu$. A rapid ``population
switching''\cite{YG04,Sindel} (which may be accompanied by the
switching of positions of the two mean-field energy levels,
$E_{1,2}$), takes place. The available results, however, remain
incomplete in that (i) the behavior of $n_i$ near switching
(abrupt vs. continuous for different values of parameters) was not
investigated, (ii) only the case of $|a_L|=|a_R|$ and
$|b_L|=|b_R|$ was considered, omitting the
important effects of coupling asymmetry (see, however, Ref. 
\onlinecite{Marquardt}), 
and (iii) the
relationship between population switching and  PLs was not
addressed fully and correctly. The present article is aimed, in part, at 
clarifying these issues.

We find that at sufficiently large $U$, including the dot-lead
coupling asymmetry largely alleviates the ``sign problem'' as
outlined above, giving rise to a phase lapse of $\Delta
\Theta_{tr} =-\pi$ between the two Coulomb peaks for the
overwhelming part of the phase diagram at both $\sigma=1$ and
$\sigma=-1$. This is a result of an effective renormalization of
the coupling sign, $\sigma=-1$ to $\sigma=1$, due to the
interaction. As some asymmetry of individual level coupling is
generally expected in  experimental realizations of QDs, this
novel phase-lapse mechanism appears relevant for understanding the
experimental data. Furthermore, we consider the
implications of interaction-induced ``population switching''\cite{SI,Baltin} 
for the transmission phase. We show that,
under certain conditions (``abrupt''
switching), this leads to a modification of phase-lapse value
($|\Delta \Theta_{tr}|<\pi$). Once  fluctuations (omitted in the
present mean-field treatment) are taken into account, this result
may translate into a more complex behavior in the vicinity of the
phase lapse.

The analysis of the full four-dimensional space of all values of
$a_{L,R}$ and $b_{L,R}$ proves too cumbersome and perhaps
redundant. Rather, we find it expedient to investigate a suitable
3D subspace, which is defined by a constraint,
$b_R^2-b_L^2=a_L^2-a_R^2$. Then there exists a unitary
transformation of the two dot operators, $\hat{d}_{1,2}
\rightarrow \tilde{d}_{1,2}$, changing  the coefficients in Eq.
(\ref{eq:tunnel}) in such a way that
$\tilde{a}_L=\tilde{a}_R\equiv a$, $\tilde{b}_L = \tilde{\sigma}
\tilde{b}_R\equiv b$ with $\tilde{\sigma}=-1$ (the $\tilde{\sigma}=1$ case
corresponds to the same-sign symmetric original coupling:
$a_L=a_R$, $b_L=b_R$). The transformation also affects the form of
the first two terms on the r.\ h.\ s. of Eq. (\ref{eq:Hdotgen}),
which now read
\begin{equation}
(\tilde{E}_1^{(0)}-\mu ) \tilde{d}^\dagger_1 \tilde{d}_1 +
(\tilde{E}_2^{(0)}-\mu ) \tilde{d}^\dagger_2 \tilde{d}_2-
\frac{w_0}{2} ( \tilde{d}^\dagger_1 \tilde{d}_2 + \tilde{d}^\dagger_2
\tilde{d}_1 ).
\label{eq:intradot}
\end{equation}
The coefficients $\tilde{E}_{1,2}^{(0)}$ and $w_0$ can be formally
thought of as the bare ``site energies'' and ``intra-dot hopping''
of a QD depicted in Fig. \ref{fig:schemedot}, and are related to
the level energies [cf. Eq. (\ref{eq:Hdotgen})] by
$2{E}_{1,2}^{(0)}= (\tilde{E}^{(0)}_1+\tilde{E}^{(0)}_2) \mp
[(\tilde{E}^{(0)}_1-\tilde{E}^{(0)}_2)^2+{w_0}^2]^{1/2}$. Our
analysis will be carried out in terms of this new QD with
$\tilde{\sigma}=\pm1$. For the $\tilde{\sigma}=-1$ case, $w_0$ is actually  a
measure of (left-right) asymmetry in the coupling of the original
QD levels, ${E}_{1,2}^{(0)}$, to the two leads.

Our calculation consists of the following steps:(i) mean field
decoupling of the interaction term in \ref{eq:Hdotgen}; (ii)
obtaining an effective single particle Hamiltonian in terms of the
averages $\langle \tilde{d}^{\dagger}_i \tilde{d}_j \rangle$;
(iii) expressing $\Theta_{tr}$ in terms of  the parameters of that
Hamiltonian; (iv) expressing $\langle \tilde{d}^{\dagger}_i
\tilde{d}_j \rangle$ in terms of $\Theta_{tr}$ employing the
Lifshits--Krein trace formalism; (v) solving self-consistently the
resultant equations for $\langle \tilde{d}^{\dagger}_i \tilde{d}_j
\rangle$ ; (vi) obtaining explicit results for $\Theta_{tr}$.

(i) {\it Mean field decoupling} reads
\begin{eqnarray}
\tilde{d}^\dagger_1 \tilde{d}^\dagger_2 \tilde{d}_2 \tilde{d}_1
&\!\rightarrow &\!\tilde{d}^\dagger_1 \tilde{d}_1 \langle
\tilde{d}^\dagger_2 \tilde{d}_2 \rangle + \tilde{d}^\dagger_2
\tilde{d}_2 \langle
\tilde{d}^\dagger_1 \tilde{d}_1 \rangle - \langle \tilde{d}^\dagger_1
\tilde{d}_1 \rangle \langle
\tilde{d}^\dagger_2 \tilde{d}_2 \rangle- \nonumber \\
&&\!-\tilde{d}^\dagger_1 \tilde{d}_2 \langle \tilde{d}^\dagger_2
\tilde{d}_1 \rangle-
\tilde{d}^\dagger_2 \tilde{d}_1 \langle \tilde{d}^\dagger_1
\tilde{d}_2 \rangle +
|\langle \tilde{d}^\dagger_1 \tilde{d}_2 \rangle|^2\,.
\label{eq:HFdecomp}
\end{eqnarray}
We verified that the results of our mean-field scheme are
independent on the basis (of the two dot states) in which the
decoupling is carried out. In the case of asymmetric coupling, it
is important\cite{Khomskii} to keep the off-diagonal ("excitonic") 
average values
in the above expression, e.g. $\langle \tilde{d}^\dagger_2
\tilde{d}_1 \rangle$. Owing to a cancellation between virtual
hopping paths between the two QD sites, these averages vanish  in
the $\tilde{\sigma}=-1$ symmetric case of $w_0=0$ (corresponding
to $a_L=a_R$, $b_L=-b_R$)  \cite{YG04,Sindel}. However, this does
not occur generally, nor indeed in the same-sign symmetric case,
leading to difficulties noted in Ref. \onlinecite{Sindel}.

(ii) Substituting  Eqs. (\ref{eq:intradot}--\ref{eq:HFdecomp})
into (\ref{eq:Hdotgen}) is tantamount to  mapping of the original
model onto an {\it effective non-interacting model} with the
Hamiltonian given by Eq. (\ref{eq:tunnel}) and the mean-field dot
term,
\begin{equation}
{\cal H}_d^{MF} = (\tilde{E}_1-\mu) \tilde{d}^\dagger_1 \tilde{d}_1 +
(\tilde{E}_2-\mu)
\tilde{d}^\dagger_2 \tilde{d}_2
-\frac{w}{2}
(\tilde{d}^\dagger_1 \tilde{d}_2 + \tilde{d}^\dagger_2 \tilde{d}_1).
\label{eq:hammf}
\end{equation}
The self-consistency conditions take the form of three coupled mean-field
equations,
\begin{eqnarray}
\tilde{E}_1 = \tilde{E}_1^{(0)} +&U& \langle \tilde{d}_2^\dagger
\tilde{d} _2 \rangle ,
\,\,\,\,\,\,
\tilde{E}_2 = \tilde{E}_2^{(0)} + U \langle \tilde{d}_1^\dagger
\tilde{d}_1 \rangle ,
\label{eq:mfee2} \\
w &=&w_0 + 2 U \langle{\tilde{d}^\dagger_1 \tilde{d}_2} \rangle.
\label{eq:mfew}
\end{eqnarray}

(iii) For  the effective single-particle   model (\ref{eq:hammf})
one can readily 
{\it compute the transmission phase},
$\Theta_{tr}(\epsilon)$. In the  $\tilde{\sigma}=-1$ case, it is given by
\begin{equation}
\sqrt{t^2-\epsilon^2}\,{\rm tan} \Theta_{tr}
=\epsilon+\frac{b^2(\tilde{E}_1-\epsilon)+
a^2(\tilde{E}_2-\epsilon)+2\epsilon \frac{a^2b^2}{t^2}}
{(\tilde{E}_1-\epsilon)(\tilde{E}_2-\epsilon)-
\frac{1}{4}w^2-\frac{a^2b^2}{t^2}}\,
\label{eq:phaseint}
\end{equation} 
(where $2t$ is the width of conduction band in the leads)
and suffers a lapse of $-\pi$ at that value of $\epsilon$ for which the
transmission vanishes, {\it i. e.}, $\epsilon=Z$,
\begin{equation} 
Z= \frac{\tilde{E}_2 a^2-\tilde{E}_1 b^2}{a^2-b^2}\,.
\label{eq:intdotzero}
\end{equation}

(iv) The quantum mechanical {\it average values} in Eqs.
(\ref{eq:mfee2}--\ref{eq:mfew}) are given by  derivatives
\begin{equation}
\langle \tilde{d}^\dagger_{1,2} \tilde{d}_{1,2} \rangle =
{\partial \Omega_{MF}}/
{\partial \tilde{E}_{1,2}}\,, \,\,\,\, \langle \tilde{d}^\dagger_1 \tilde{d}_2
\rangle = -{\partial \Omega_{MF}}/{\partial w} \,
\end{equation}
of the thermodynamic potential of the effective system. The latter
is evaluated exactly with the help of  the Lifshits--Krein trace
formula\cite{iml},
\begin{equation}
\Omega_{MF}=\Omega_0+ \int_{-t}^\mu \xi(\epsilon) d\epsilon\,.
\label{eq:trace}
\end{equation}
Here, $\Omega_0$ is the combined potential of a disconnected
system comprising a  dot [Eq. (\ref{eq:hammf})] and a wire,
\begin{equation}
{\cal H}_w= -\frac{t}{2} \sum_j \left(\hat{c}^\dagger_j
\hat{c}_{j+1} + \hat{c}^\dagger_{j+1} \hat{c}_j \right)- \mu
\sum_j \hat{c}^\dagger_j \hat{c}_j. \label{eq:wire}
\end{equation}
 The spectral shift function $\xi$ is defined by its
relationship, 
\begin{equation}
d \xi/d\epsilon=-\delta \nu(\epsilon)\,, 
\label{eq:defxi}
\end{equation}
to the
change of the total density of states of this system  due to
 a local perturbation,
\begin{equation}
V=V_T+\frac{t}{2}\left(\hat{c}^\dagger_{1/2}\hat{c}_{-1/2}+
\hat{c}^\dagger_{-1/2}\hat{c}_{1/2}  \right).
 \label{eq:tunnel2}
\end{equation}
\begin{figure*}
\includegraphics{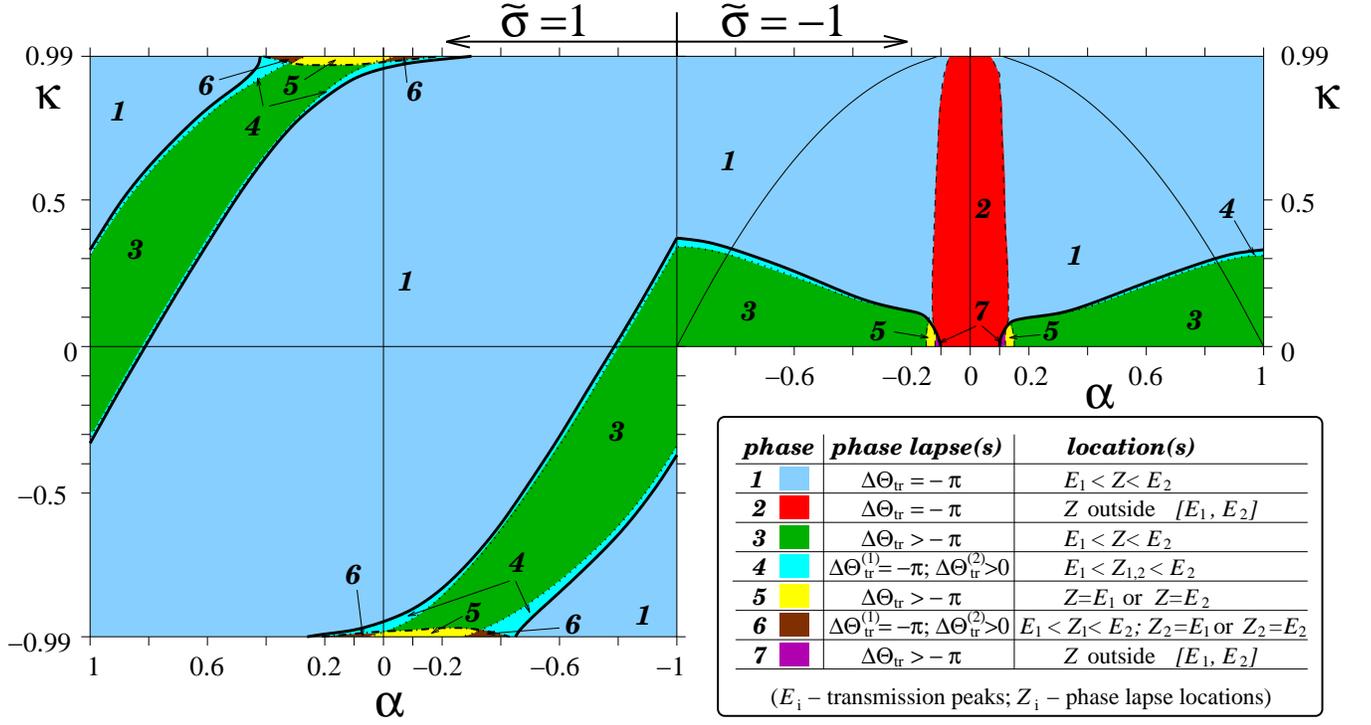}
\caption{\label{fig:phadiag} (color)
The ``phase diagram'' of a two-level QD with
$\tilde{a}_L=\tilde{a}_R= a$, $\tilde{b}_L = \tilde{\sigma}
\tilde{b}_R= b$.
The parameters are $U=0.1t$, $\tilde{E}_1^{(0)}=0$, $\tilde{E}_2^{(0)}=0.004t$,
and $\sqrt{a^2+b^2}=0.125t$.
The axes represent
the 1-2 level asymmetry, $\alpha = (|a|-|b|)/\sqrt{a^2+b^2}$,
and the dimensionless intra-dot hopping, $\kappa=w_0/
[(\tilde{E}_1^{(0)}-\tilde{E}_2^{(0)})^2+w_0^2]^{1/2}$.
Properties of different phases are summarized in the table. At $U
\leq 0.04t$, the border of discontinuous-evolution region (bold
line) does not meet the boundary between phases 1 and 2.}


\end{figure*}
\begin{figure}
\includegraphics{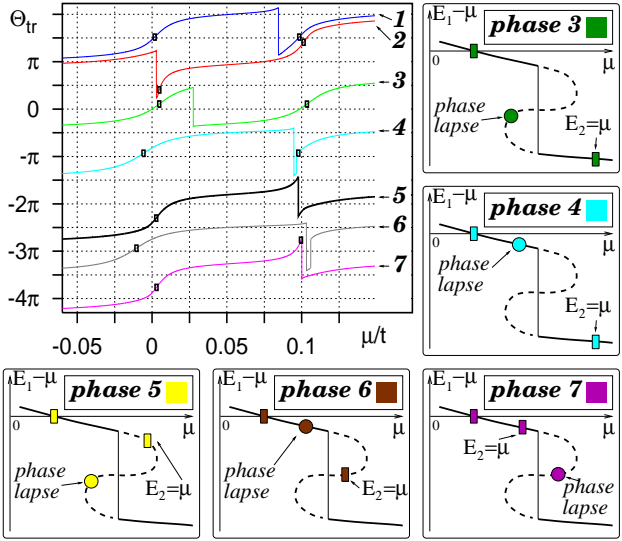}
\caption{\label{fig:plot} (color) Typical behavior of
$\Theta_{tr}(\mu)$ in different phases (top left; plots shifted
for convenience). Relative positions of transmission peaks (boxes; also in the
main panel)
and the $-\pi$-PLs (circles) in phases 3-7 are clarified by the
schematic plots of $E_1-\mu$ around the multiple-solution
region (absent for phases 1-2). Solid (dashed) lines correspond
to stable (unstable) solutions. The abrupt ``switching'' of
solutions (vertical solid line) may either renormalize the PL
 (when the $- \pi$-lapse lies in the unstable region) or result in a
positive jump of $\Theta_{tr}$.}
\end{figure}
In Eq. (\ref{eq:tunnel2}), the second term on the r.h. s.
corresponds to cutting the link between  sites $j=-1/2$ and
$j=1/2$ of the wire. The two resulting leads are  coupled to the
QD by  $V_T$ [Eq. (\ref{eq:tunnel})].
 Since for a wire of a finite length $2L$, $\xi$ is related to
the shifts of (discrete) energy levels under the effect of $V$, it
is easy \cite{Anderson} to express $\xi$ in terms of
$\Theta_{tr}$, {\it viz.} 
\begin{equation}
\xi=-\Theta_{tr}/\pi+m(\epsilon)
\label{eq:friedel}
\end{equation}
(see Appendix \ref{app:shift}). The
integer-valued function $m$ should be chosen to satisfy the
requirement\cite{iml} for $\xi(\epsilon)$ to vanish continuously
with decreasing strength of the perturbation ({\it e.g.,} $\lambda
V$ with $\lambda \rightarrow 0$). We find that the value of $m$
changes by $+1$ at  $\epsilon=E_{1,2}$ [eigenvalues of ${\cal
H}^{MF}_d$, Eq.(\ref{eq:hammf})], and by $-1$ at the transmission
zero. For $\tilde{\sigma}=-1$ we obtain (in the units where $t=1$):
\begin{eqnarray}
\langle \tilde{d}^\dagger_1 \tilde{d}_1 \rangle
 &=& \int_{-1}^\mu \sqrt{1-\epsilon^2}
\,X(\epsilon) d \epsilon \times
\label{eq:oppmfee1}\\
&&\!\!\!\times
\left[a^2b^4(1-\epsilon^2)+a^2 (\tilde{E}_2-\epsilon+b^2\epsilon)^2+
\frac{b^2w^2}{4}
\right] \,, \nonumber\\
\langle \tilde{d}^\dagger_2 \tilde{d}_2
\rangle& =& \int_{-1}^\mu \sqrt{1-\epsilon^2}
\,X(\epsilon) d \epsilon \times
\label{eq:oppmfee2}\\
&&\!\!\!\times
\left[a^4b^2(1-\epsilon^2)+b^2 (\tilde{E}_1-\epsilon+a^2\epsilon)^2+
\frac{a^2w^2}{4}
\right] \,, \nonumber\\
\langle \tilde{d}^\dagger_1 \tilde{d}_2 \rangle& =&\frac{w}{2}
\int_{-1}^\mu \sqrt{1-\epsilon^2}
\,X(\epsilon) d \epsilon \times
\label{eq:oppmfew}\\
&&\!\!\!\times \left[a^2 (\tilde{E}_2-\epsilon) +b^2 (\tilde{E}_1-
\epsilon) + 2a^2b^2 \epsilon \right] \, , \nonumber
\end{eqnarray}
\begin{eqnarray}
\frac{1}{\pi X(\epsilon)}=(1-\epsilon^2)\left[a^2(\tilde{E}_2-\epsilon)
+b^2 (\tilde{E}_1-\epsilon) +2a^2b^2\epsilon \right]^2\!\!+ \nonumber \\
+\!\left[(\tilde{E}_1-\epsilon+a^2\epsilon)(\tilde{E}_2-\epsilon+b^2\epsilon)
\!-\!(1-\epsilon^2)a^2b^2\!-\frac{1}{4}w^2\right]^2\!\!. \nonumber
\end{eqnarray}
Similar expressions are obtained also for the $\tilde{\sigma}=1$ case
(see Appendix \ref{app:samesign}).

We now {\it solve} equations (\ref{eq:mfee2}--\ref{eq:mfew})
numerically (v),
and substitute the resulting
values of $\tilde{E}_{1,2}$ and $w$ into the expression for 
$\Theta_{tr}$ to get the {\it transmission phase} (vi).

The results are summarized in the phase diagram, Fig.
\ref{fig:phadiag}.
The left-hand part corresponds to  $\tilde{\sigma}=1$, whereas the
$\tilde{\sigma}=-1$ case (when the results do not depend on the
sign of $w_0$) is shown on the right. The bold line marks the
boundary between continuous (phases 1-2) and discontinuous (see
below) regimes of dependence of the effective QD parameters on
$\mu$. Within each regime, different phases are identified
according to the magnitude and location of PL(s) with respect to
the transmission peaks (Fig. \ref{fig:phadiag}, table). It should
be noted that in the $\tilde{\sigma}=-1$ case the latter are given
by $\mu_{1,2} = (\tilde{E}_1+\tilde{E}_2)/2 \mp \frac{1}{2}
[(\tilde{E}_1-\tilde{E}_2)^2+w^2+4a^2b^2/t^2]^{1/2}$, and are
slightly shifted with respect to mean field dot levels, $E_{1,2}$.
In the table, we denote transmission peaks by $E_{1,2}$
irrespective of the sign of $\tilde{\sigma}$ in order to keep the
notation uniform. Typical dependence of $\Theta_{tr}$ on $\mu$ for
each phase is shown in Fig. \ref{fig:plot}.

In the continuous-evolution part, phase 1 (phase 2), which
occupies a large (small) area  of the  phase space, corresponds to
the case when the phase lapse of $-\pi$, associated with the
transmission zero, lies within (outside) the interval of values of
$\mu$ between the two transmission peaks. It should be noted that
the right-hand, $\tilde{\sigma}=-1$, side is expected to be
representative of both opposite- and same-sign cases ($\sigma=\pm
1$), provided that the left-right asymmetry is sufficiently strong
(large $w_0$). This is illustrated by the thin solid line, above
(below)  which coupling signs for the two bare dot levels
$E_{1,2}^{(0)}$ become the same, $\sigma=1$ (opposite,
$\sigma=-1$). Once the interaction effects are taken into account,
one sees that phase 1 extends also far below this line, which is
indicative of the effective change of the coupling sign [due in
turn to the interaction-induced enhancement of $w$; at $w \gg \tilde{E}_2-
\tilde{E}_1$, the coupling of the two mean-field dot levels, $(\tilde{d}_1 \pm
\tilde{d}_2)/\sqrt{2}$, to the leads is same-sign].

 The discontinuous behavior
is associated with the presence of multiple solutions of the mean
field equations (\ref{eq:mfee2}--\ref{eq:mfew}) within a range of
values of $\mu$, which is illustrated by a ``fold'' (bold solid
and dashed lines) on the schematic $E_1(\mu)-\mu$ plots in Fig.
\ref{fig:plot}. We find that if a system formally is allowed to
follow such a multiple-valued solution from left to right, the
value of $\Theta_{tr}$ increases, and also suffers a PL of
$-\pi$ at some point (marked by a circle). In reality,
thermodynamics dictates that the full thermodynamic potential
$\Omega=\Omega_{MF}-U\langle \tilde{d}^\dagger_1 \tilde{d}_1 \rangle \langle
\tilde{d}_2^\dagger \tilde{d}_2\rangle +U \langle \tilde{d}_1^\dagger 
\tilde{d}_2\rangle^2$ [cf.
Eq. (\ref{eq:trace})] should be minimized to identify the stable
solution, resulting in a ``jump'' (vertical line), which in turn
is associated with a positive increase of $\Theta_{tr}$ by a
fraction of $\pi$, giving rise to a second ``PL'' (phases 4,6), and
with the population switching\cite{SI,Baltin} of the dot levels. If
the transmission zero lies in the thermodynamically unstable part
of the solution (bold dashed line; phases 3,5,7), the PL of $-\pi$
should be added to this positive increase of $\Theta_{tr}$, giving
rise to a single ``renormalized'' PL. Finally, one of the
transmission peaks may be located within the unstable region
(phases 5,6) with a result that the plot of $\Theta_{tr}(\mu)$
does not have a corresponding inflection point, which is replaced
by a PL.

It follows that at least within the mean-field framework  discontinuous
population switching is always associated with the presence of multiple 
solutions and hence with ``renormalized'' PLs (or alternatively with
additional ``PLs'' characterized by an increase of phase by a fraction of 
$\pi$). This conclusion is clearly at variance with the suggestion of Ref.
\onlinecite{SI} that the discontinuous switching between multiple solutions
gives rise to the PLs of $\pi$ as observed experimentally. We note that while
the behavior of transmission phase in this regime should be investigated
beyond the mean field, the main point of our paper is that there is another
mechanism which gives rise to a PL of $\pi$ {\it without} a discontinuous
population switching (phase 1). Since this latter scenario does not involve
instabilities of any kind, it can be expected to remain  
robust with respect to fluctuations (not included in the present treatment). 

In summary, we have presented here  a  generic mechanism for the
appearance of  phase lapses between Coulomb blockade peaks. These
PLs may be renormalized by a discontinuous  "population
switching". Experimentally it would be interesting to correlate
the latter with the former by simultaneously measuring  dot
occupancy (employing a quantum point contact), and transmission
phase. Theoretically, going beyond a  mean field analysis is
needed to determine the importance of quantum fluctuations.

The authors  thank R. Berkovits, M. Heiblum, I. V. Lerner, V. Meden, and Y.
Oreg for enlightening discussions. This work was  supported by the
ISF (grant No. 193/02-1 and the Centers of Excellence Program), by
the EC RTN Spintronics, the BSF (grants \# 2004162 and \# 703296),
and by the Israeli Ministry of Absorption. YG was also supported by an
EPSRC fellowship.

\appendix
\section{Spectral Shift Function, Transmission Phase, and Friedel Sum Rule}
\label{app:shift}

For the case at hand, the use of the standard formula\cite{iml} for the 
spectral shift function $\xi$,
\begin{equation}
\xi(\epsilon)= -\frac{1}{\pi} {\rm Arg} \, {\rm Det} \left\{
\hat{1} - \frac{1}{\epsilon - {\rm i0}-{\cal H}_w -{\cal H}_d^{MF}} 
\hat{V} \right\}\,,
\label{eq:lkspectral}
\end{equation}  
proves rather cumbersome. Instead, we will use the underlying notion 
of spectral shifts\cite{iml} in order to derive the generic 
relation (\ref{eq:friedel}) 
between $\xi$ and the transmission phase. This derivation also allows for an 
important insight concerning the Friedel sum rule.  

We consider a system similar to that shown in Fig. \ref{fig:schemedot}, with 
the QD between the sites $-1/2$ and $1/2$ replaced by an arbitrary  point 
scatterer. The latter is characterized by an $S$-matrix whose elements have 
a smooth dependence on the particle energy\cite{blackbox}. While the 
boundary conditions cannot affect the value of
$\xi$ in the limit when the length of the wire, $2L$, is large, the treatment
is simpler when periodic boundary conditions are assumed. The spectrum 
of the wire in the absence of the scatterer, which we 
refer to as {\it unperturbed}, is then given by
\begin{equation}
\epsilon(k_j)=-\cos k_j\,,\,\,\,\,k_j=\frac{\pi j}{L}\,,\,\,\,j=0,1,2,...,L
\label{eq:unperturbed}
\end{equation}
[cf. Eq. (\ref{eq:wire}) where we assumed $t=1$].
The wave functions are proportional to 
$\exp{(\pm {\rm i}k_jx)}$ and, for $j \neq 0,L$, the corresponding energy 
levels are  doubly
degenerate. Since we are ultimately interested in the $L\rightarrow \infty$
limit, it is assumed that the inter-level spacing in the wire constitutes
the smallest energy scale in the problem. The levels are shifted, and the 
degeneracy is lifted, in the presence of the scatterer, when the wave function
is generally given by
\begin{equation}
\psi(x)=\left\{\begin{array}{ll} A_1 {\rm e}^{{\rm i}kx}+ B_1 
{\rm e}^{-{\rm i}kx}, & x<0,  \\
A_2 {\rm e}^{{\rm i}kx}+ B_2 
{\rm e}^{-{\rm i}kx}, & x>0.  \end{array} \right.
\label{eq:wavefunction}
\end{equation}
The linear relationship between coefficients on the right and on 
the left of the scatterer reads (assuming time-reversal symmetry)\cite{LL}:
\begin{equation}
A_2=\alpha A_1 + \beta B_1\,,\,\,\,B_2=\beta^* A_1+ \alpha^* B_1
\end{equation}
with $|\alpha|^2-|\beta|^2=1$. Relation of the quantities $\alpha$ and $\beta$ 
to the $S$-matrix is given by, 
{\it e.g.},
setting $A_1=0$ (incoming particle from the right), hence (right-right) 
reflection amplitude, $r_{rr}=\beta/\alpha^*$ and  transmission amplitude, 
$t_{tr}=1/\alpha^*$. 

Now the periodic boundary conditions dictate that the allowed momentum values
shift,
\begin{equation}
k_j \rightarrow k^{(1,2)}_j = k_j+\frac{\Delta^{(1,2)}_j  \pi}{L},
\label{eq:shiftk}
\end{equation}   
Substituting Eqs. (\ref{eq:wavefunction}--\ref{eq:shiftk}) into the condition
$\psi(-2L+0) = \psi(+0)$, we find for $l=1,2$
\begin{eqnarray}
A_1 {\rm e}^{- 2 \pi {\rm i} \Delta^{(l)}}&=& A_2=\alpha A_1 + \beta B_1 ,\nonumber\\
B_1 {\rm e}^{2 \pi {\rm i} \Delta^{(l)}}&=& B_2=\beta^* A_1+ \alpha^* B_1. \nonumber
\end{eqnarray}
This yields the equation for $\Delta_j^{(1,2)}$ (cf. Ref. \onlinecite{Anderson}):
\begin{equation}
\alpha {\rm e}^{4 \pi {\rm i} \Delta^{(l)}}- 2{\rm e}^{2 \pi {\rm i} \Delta^{(l)}}
+ \alpha^*=0, 
\end{equation}
or equivalently
\begin{eqnarray}
\left({\rm e}^{2 \pi {\rm i} \Delta^{(l)}}-{\rm e}^{2 \pi {\rm i} \Delta^{(1)}_j}
\right)  
\left({\rm e}^{2 \pi {\rm i} \Delta^{(l)}}-{\rm e}^{2 \pi {\rm i} \Delta^{(2)}_j}
\right)= \nonumber \\
={\rm e}^{4 \pi {\rm i} \Delta^{(l)}}- \frac{2}{\alpha}{\rm e}^{2 \pi {\rm i} \Delta^{(l)}}
+ \frac{\alpha^*}{\alpha},
\nonumber
\end{eqnarray}
yielding
\begin{equation}
{\rm e}^{2 \pi {\rm i} (\Delta^{(1)}_j+\Delta^{(2)}_j)}
=\frac{\alpha^*}{\alpha}\,.
\end{equation}
In the limit $L\rightarrow
\infty$, the quantities $\Delta^{(1,2)}_j$ become functions of energy and,
writing also $\alpha = {\rm exp}({{\rm i} \Theta_{tr}})/|t_{tr}|$ with 
$\Theta_{tr}$ the
transmission phase, we
find
\begin{equation}
\Delta[\epsilon (k_j)] \equiv \Delta^{(1)}_j+\Delta^{(2)}_j = 
-\frac{1}{\pi}\Theta_{tr}[\epsilon(k_j)] + 
m[\epsilon(k_j)]\,.
\label{eq:Delta2}
\end{equation} 

Let us now discuss the quantities appearing in Eq. (\ref{eq:Delta2}). (i)
$\Theta_{tr}$ is the transmission phase.  In the
presence of localized states within the scatterer (dot levels $E^d_i$), 
$\Theta_{tr}$ increases by $\pi$ as the energy of interest [$\epsilon (k_j)$
in our notation, or more physically, the chemical potential] spans a
resonance. (ii) $m(\epsilon)$ is an integer which we will now choose in
such a way that $\Delta$ coincides with the Lifshits -- Krein spectral shift
function, $\xi(\epsilon)$ [see Eq. (\ref{eq:defxi})]. $m(\epsilon)$ then changes by $+1$ with increasing energy at every $E^d_i$; in addition, it changes 
by $-1$ at the points where transmission vanishes (transmission PLs). 
We thus arrive at Eq. (\ref{eq:friedel}). (iii)
$\Delta^{(1)}_j \pi/L$ and $\Delta^{(2)}_j \pi/L$ are the shifts in the allowed
values of momentum [cf Eq. (\ref{eq:shiftk})]. 

There is no bound state corresponding to a PL, implying that $\xi(\epsilon)$
should be continuous at that point (transmission zero). The choice of 
$m(\epsilon)$ discussed above [along with Eq. (\ref{eq:Delta2})], ensures that
$\Delta$ indeed may be synonymous with $\xi$ (see below).

In order to use the calculated value of spectral shift function for the 
total energy evaluation via  the trace formula, 
Eq. (\ref{eq:trace}), one needs to know the overall additive constant in
$\xi(\epsilon)$. In the regime of interest to us, no bound state is formed
below the band bottom (at $\epsilon < -1$). From the viewpoint of  the 
lowest-energy electron states ($\epsilon\rightarrow -1+0$), 
the scatterer then acts as an impenetrable potential barrier
(and not as a potential well), and the constant is fixed by a readily
derivable condition, $\xi(\epsilon \rightarrow -1+0)=1/2$, valid for any 
barrier with no bound state formed below its bottom.

The quantity $\xi(\epsilon)$ remains a smooth function of 
energy $\epsilon$ away from band edges $\epsilon = \pm 1$ and the dot 
levels ${E}^d_i$. As mentioned in the text, the spectral shift function 
is related to the perturbation-induced change
in the density of states. For the unperturbed system, the latter can be defined
{\it in the} $L\rightarrow \infty$ {\it limit only} as 
\begin{equation}
\nu_0[\epsilon(k_j)] = \lim_{L\rightarrow\infty} \frac{2}{\epsilon(k_{j+1}) -
\epsilon(k_j)} \propto L\,,
\end{equation}
where the factor of $2$ reflects the double degeneracy of energy levels. In the
presence of the scatterer we obtain, with the help 
of Eq. (\ref{eq:shiftk}),
\begin{eqnarray}
&&\nu[\epsilon(k_j)]=\nonumber \\
&&=\lim_{L\rightarrow\infty} 
\frac{2}{\left[\epsilon(k_{j+1})
 -\epsilon(k_j) \right] \left\{1+\frac{1}{2}\xi[\epsilon(k_{j+1})]-
\frac{1}{2}\xi[\epsilon(k_{j})
]\right\}} \nonumber \\
&&=\nu_0[\epsilon(k_j)]+ \delta \nu[\epsilon(k_j)]\,,\,\,\,\,\nonumber
\end{eqnarray}
where
\begin{equation}
\delta\nu(\epsilon) = - d \xi(\epsilon) /d \epsilon\,.
\label{eq:deltanu}
\end{equation}
Here, we used the obvious fact that the centre of gravity of the two perturbed
levels formed out of a doubly degenerate unperturbed level 
$\epsilon(k_j)$ is given
by $\epsilon(k_j)+ \frac{1}{2} [\epsilon(k_{j+1})-\epsilon(k_j)] 
\Delta[\epsilon(k_j)]$ (substituting $\Delta$ with $\xi$ for our choice of 
$m$). We note that, as expected on physical grounds, the
quantity $\delta \nu$ is not extensive, {\it i. e.}, it is not proportional
to the length of the wire (in contrast to $\nu_0$). For the specified choice 
of $m(\epsilon)$ in Eq. (\ref{eq:Delta2}), Eq. (\ref{eq:deltanu}) yields
also a delta-functional contribution to $\delta \nu$ of the form 
$-\sum_i \delta(\epsilon-{E}^d_i)$. This corresponds to merging of the
discrete dot levels into continuum and shows that $\delta \nu$ is the 
difference in the density of states between the wire with the scatterer and  
a disconnected system comprised of an unperturbed wire alongside an 
isolated scatterer.

Integrating Eq. (\ref{eq:deltanu}), we get the expression for 
the total particle number,
\begin{equation}
N(\mu)=N^{(0)}_{wire}(\mu)+\sum_i \theta(\mu-{E}_i^d)-\xi(\mu)\,,
\label{eq:friedel2}
\end{equation}
where the first term on the r.\ h.\ s. is the band filling of an unperturbed
wire. By re-writing this in terms of transmission phase 
$\Theta_{tr}(\epsilon)$ [cf.
Eq. (\ref{eq:Delta2})], we get the {\it Friedel sum rule} in the form
\begin{equation}
N(\mu)=N^{(0)}_{wire}(\mu)+\Theta_{tr}(\mu)/\pi+ \tilde{m} (\mu)
\label{eq:friedel3}
\end{equation}
With increasing $\mu$, the integer $\tilde{m}$ changes by $+1$ at transmission
zeroes, $\mu=Z_i$. We note that the sum of the two last terms on
the r.\ h.\ s. of Eq. (\ref{eq:friedel3}) remains continuous at $\mu=Z_i$,
emphasizing that the Friedel sum rule does {\it not} account for the 
transmission phase lapses. This is because the underlying spectral 
characteristic, $\xi(\epsilon)$  [cf. Eq. (\ref{eq:friedel2})] remains smooth 
at $\mu=Z_i$ and in general does not depend on $|t_{tr}|$.

\section{Mean-Field Equations in the \protect ${\mathbf {\tilde{\sigma}=1}}$ Case}
\label{app:samesign}

In the case of same-sign symmetric coupling of the QD to the leads, $\tilde{\sigma}=1$,
Eqs. (\ref{eq:phaseint}--\ref{eq:intdotzero}) and   
(\ref{eq:oppmfee1}--\ref{eq:oppmfew}) are replaced with
\begin{equation}
\sqrt{t^2-\epsilon^2}\,{\rm tan} \Theta_{tr}=
\epsilon+\frac{b^2(\tilde{E}_1-\epsilon)+a^2(\tilde{E}_2-\epsilon)+abw}
{(\tilde{E}_1-\epsilon)(\tilde{E}_2-\epsilon)-
\frac{1}{4}w^2}\,,\nonumber
\end{equation} 
\begin{equation}
Z= \frac{\tilde{E}_2 a^2+\tilde{E}_1 b^2+ abw}{a^2+b^2}\, \nonumber
\end{equation}
and
\begin{eqnarray}
\langle \tilde{d}^\dagger_1 \tilde{d}_1 \rangle 
&=& \int_{-1}^\mu \sqrt{1-\epsilon^2}
\left[a(\tilde{E}_2-\epsilon) + \frac{1}{2}bw \right]^2 
Y(\epsilon) d \epsilon  \,,
\nonumber\\
\langle \tilde{d}^\dagger_1 \tilde{d}_1 \rangle 
&=& \int_{-1}^\mu \sqrt{1-\epsilon^2}
\left[b(\tilde{E}_1-\epsilon) + \frac{1}{2}aw \right]^2 Y(\epsilon)
 d \epsilon  \,,
\nonumber\\
\langle \tilde{d}^\dagger_1 \tilde{d}_2 \rangle 
&=& \int_{-1}^\mu \sqrt{1-\epsilon^2}
\,Y(\epsilon) d \epsilon \times 
\nonumber\\
&&\times 
\left[a(\tilde{E}_2-\epsilon) + \frac{1}{2}bw \right] \cdot
\left[b(\tilde{E}_1-\epsilon) + \frac{1}{2}aw \right]\,,\nonumber
\end{eqnarray}
respectively. Here,
\begin{eqnarray}
\frac{1}{\pi Y(\epsilon)}&=&(1-\epsilon^2)\left[(\tilde{E}_1-\epsilon)
(\tilde{E}_2-\epsilon) - \frac{1}{4}w^2 \right]^2+ \nonumber \\
&&+\left\{ a^2(\tilde{E}_2-\epsilon)+b^2(\tilde{E}_1-\epsilon)+ abw+ 
\begin{array}{c}\,\\ \,\end{array}\right. 
\nonumber \\
&&+ \left. \epsilon \left[(\tilde{E}_1-\epsilon)
(\tilde{E}_2-\epsilon) - \frac{1}{4}w^2 \right] \right\}^2.
\nonumber 
\end{eqnarray}

\end{document}